\documentstyle[twoside]{article}
\newcommand{\SAk}{S_{{\cal A}k}}
\newcommand{\quartic}{g}
\newcommand{\geff}{\lambda}

\newcount\Mac  \Mac=0  
\newcommand{\ifMac}[2]{\ifnum\Mac=1 #1 \else #2 \fi}

\newcommand{\One}{\hbox{1\kern-.24em I}}

\newcommand{\eq}[1]{~{\rm (\ref{eq:#1})}}

\def\putps(#1,#2)(#3,#4)#5#6{\put(#1,#2){\special{picture #5}}\put(#3,#4){\includegraphics{#6}}}

\def\Red{}
\def\Black{}
\def\Blue{}

\newcommand{\lascia}[1]{}
\makeatletter
\def\art{\@ifnextchar[{\eart}{\oart}}
\def\eart[#1]#2#3#4#5#6{{\rm #2}, {\em #3 \bf #4} {\rm (#6) #5}}
\def\hepart[#1]#2{{\rm #2, \em#1}}
\newcommand{\oart}[5]{{\rm #1}, {\em #2 \bf #3} {\rm (#5) #4}}

\newcounter{alphaequation}[equation]
\def\thealphaequation{\theequation\hbox to
0.6em{\hfil\alph{alphaequation}\hfil}}
\def\eqnsystem#1{
\def\@eqnnum{{\rm (\thealphaequation)}}
\def\@@eqncr{\let\@tempa\relax \ifcase\@eqcnt \def\@tempa{& & &} \or
  \def\@tempa{& &}\or \def\@tempa{&}\fi\@tempa
  \if@eqnsw\@eqnnum\refstepcounter{alphaequation}\fi
\global\@eqnswtrue\global\@eqcnt=0\cr}
\refstepcounter{equation} \let\@currentlabel\theequation \def\@tempb{#1}
\ifx\@tempb\empty\else\label{#1}\fi
\refstepcounter{alphaequation}
\let\@currentlabel\thealphaequation
\global\@eqnswtrue\global\@eqcnt=0 \tabskip\@centering\let\\=\@eqncr
$$\halign to \displaywidth\bgroup \@eqnsel\hskip\@centering
$\displaystyle\tabskip\z@{##}$&\global\@eqcnt\@ne
\hskip2\arraycolsep\hfil${##}$\hfil& \global\@eqcnt\tw@\hskip2\arraycolsep
$\displaystyle\tabskip\z@{##}$\hfil
\tabskip\@centering&\llap{##}\tabskip\z@\cr}

\def\endeqnsystem{\@@eqncr\egroup$$\global\@ignoretrue} \makeatother

\def\circa#1{\,\raise.3ex\hbox{$#1$\kern-.75em\lower1ex\hbox{$\sim$}}\,}

\newcommand{\St}{\tilde{S}}

\newcommand{\phibounce}{\phi_b}

\newcommand{\beq}{\begin{eqnarray}}
\newcommand{\eeq}{\end{eqnarray}}

\def \lta {\mathrel{\vcenter
     {\hbox{$<$}\nointerlineskip\hbox{$\sim$}}}}

\makeatletter
\def\hdashline#1(#2){\leavevmode\hbox to \z@{\baselineskip \z@%
\lineskip \z@%
\@dashdim=#2\unitlength%
\@dashcnt=\@dashdim \advance\@dashcnt 200
\@dashdim=#1\unitlength\divide\@dashcnt \@dashdim
\ifodd\@dashcnt\@dashdim=\z@%
\advance\@dashcnt \@ne \divide\@dashcnt \tw@ 
\else \divide\@dashdim \tw@ \divide\@dashcnt \tw@
\advance\@dashcnt \m@ne
\setbox\@dashbox=\hbox{\vrule \@height \@halfwidth \@depth \@halfwidth
\@width \@dashdim}\put(0,0){\copy\@dashbox}%
\put(#2,0){\hskip-\@dashdim\copy\@dashbox}%
\multiply\@dashdim 3 
\fi
\setbox\@dashbox=\hbox{\vrule \@height \@halfwidth \@depth \@halfwidth
\@width #1\unitlength\hskip #1\unitlength}\@tempcnta=0
\put(0,0){\hskip\@dashdim \@whilenum \@tempcnta <\@dashcnt
\do{\copy\@dashbox\advance\@tempcnta \@ne }}\@tempcnta=0
}\@makepicbox(#2,0)}

\def\vdashline#1(#2){\leavevmode\hbox to \z@{\baselineskip \z@%
\lineskip \z@%
\@dashdim=#2\unitlength%
\@dashcnt=\@dashdim \advance\@dashcnt 200
\@dashdim=#1\unitlength\divide\@dashcnt \@dashdim
\ifodd\@dashcnt \@dashdim=\z@%
\advance\@dashcnt \@ne \divide\@dashcnt \tw@
\else
\divide\@dashdim \tw@ \divide\@dashcnt \tw@
\advance\@dashcnt \m@ne
\setbox\@dashbox\hbox{\hskip -\@halfwidth
\vrule \@width \@wholewidth 
\@height \@dashdim}\put(0,0){\copy\@dashbox}%
\put(0,#2){\lower\@dashdim\copy\@dashbox}%
\multiply\@dashdim 3
\fi
\setbox\@dashbox\hbox{\vrule \@width \@wholewidth 
\@height #1\unitlength}\@tempcnta0
\put(0,0){\hskip -\@halfwidth \vbox{\@whilenum \@tempcnta < \@dashcnt
\do{\vskip #1\unitlength\copy\@dashbox\advance\@tempcnta \@ne }%
\vskip\@dashdim}}\@tempcnta0
}\@makepicbox(0,#2)}
\newdimen\sideftwd
\newbox\local@box\newbox\local@hbox

\def\lfigure#1{\if@twocolumn\local@ft{figure}{#1}\else\side@ft{figure}{#1}\fi}
\def\rfigure#1{\if@twocolumn\local@ft{figure}{#1}\else\side@ft{figure}{#1}\fi}
\def\ltable#1{\if@twocolumn\local@ft{table}{#1}\else\side@ft{table}{#1}\fi}
\def\rtable#1{\if@twocolumn\local@ft{table}{#1}\else\side@ft{table}{#1}\fi}
\def\endlfigure{\endlside@ft}
\def\endrfigure{\endrside@ft}
\def\endltable{\endlside@ft}
\def\endrtable{\endrside@ft}
\def\side@ft#1#2{\par
\sideftwd=#2
\def\@captype{#1}
\setbox\@tempboxa\vtop\bgroup\textwidth=\sideftwd
\columnwidth=\sideftwd \hsize\columnwidth
\@parboxrestore}
\def\endlside@ft{\egroup
\@tempdima=\ht\@tempboxa
\advance\@tempdima by \dp\@tempboxa
\@tempcnta=\@tempdima
\divide\@tempcnta by \baselineskip
\advance\@tempcnta by 2
\global\hangindent\sideftwd
\global\hangafter-\@tempcnta
\noindent \dp\@tempboxa=\z@ \ht\@tempboxa=\z@
\hbox to \z@{
\hbox to \z@{\hss\box\@tempboxa}\hss}%
\hskip\parindent
\global\@ignoretrue}
\def\endrside@ft{\egroup
\@tempdima=\ht\@tempboxa
\advance\@tempdima by \dp\@tempboxa
\@tempcnta=\@tempdima
\divide\@tempcnta by \baselineskip
\advance\@tempcnta by 2
\global\hangindent-\sideftwd
\global\hangafter-\@tempcnta
\noindent \dp\@tempboxa=\z@ \ht\@tempboxa=\z@
\hbox to \z@{\hskip\textwidth
\hbox to \z@{\hss\box\@tempboxa}\hss}%
\hskip\parindent
\global\@ignoretrue}
\def\local@ft#1{\def\@captype{#1}
\setbox\local@box\vbox\bgroup
\boxmaxdepth\z@\hsize0.9\columnwidth}
\def\endlocal@ft{\egroup
\[\hbox{\lower1ex\box\local@box}\]
\global\@ignoretrue}
\def\localfigure{\local@ft{figure}}
\def\localtable{\local@ft{table}}
\def\endlocalfigure{\endlocal@ft}
\def\endlocaltable{\endlocal@ft}
\makeatother

\begin{document}
\centerline{cond-mat/0002278 \hfill IFUP--TH/2000--01 \hfill MS--TPI/00--2}
\vspace{5mm}

\Black
\vspace{0.5cm}
\centerline{\LARGE\bf\Red Comparison of two methods}\medskip
\centerline{\LARGE\bf\Red for calculating nucleation rates}
\bigskip\bigskip\Black
\centerline{\large\bf G. M\"unster$^{\rm a}$, A. Strumia$^{\rm b}$ 
{\rm and} N. Tetradis$^{\rm c}$} 
\bigskip
\centerline{(a) \em Institut f\"ur Theoretische Physik I, Universit\"at M\"unster}
\centerline{~\em Wilhelm-Klemm-Str.~9, D-48149 M\"unster, Germany}
\smallskip
\centerline{(b) \em Dipartimento di Fisica, Universit\`a di Pisa and INFN, I-56127 Pisa, Italia}\smallskip
\centerline{(c) \em Scuola Normale Superiore, Piazza dei Cavalieri 7, I-56126 Pisa, Italy and}
\centerline{~\em Department of Physics, University of Athens, GR-15771 Athens, Greece}
\vspace{0.8cm}
\centerline{\large\bf Abstract}
\begin{quote}\large\indent
First-order phase transitions that proceed via nucleation of bubbles
are described by the homogeneous nucleation theory of Langer.
The nucleation rate is one of the most interesting parameters of these transitions.
In previous works we have computed nucleation rates with two different methods:
(${\cal A}$) employing coarse-grained potentials;
(${\cal B}$) by means of a saddle-point approximation, using dimensionally regularized field theory.
In this article we compare the results of the two approaches in order to
test their reliability and to determine the regions of applicability.
We find a very good agreement.
\end{quote}\Black
\vspace{0.5cm}

\section{Introduction}

The calculation of bubble-nucleation rates 
during first-order phase transitions is a difficult problem with
a long history. It has applications to a variety of physical situations,
ranging from the metastability of statistical systems, such as supercooled
liquids or vapours \cite{abraham}, to cosmological phase transitions,
such as the electroweak or the deconfinement phase transition in QCD 
\cite{karsch}. 
(For reviews with an extensive list of
references, see refs.~\cite{abraham,karsch,gunton,binder,rikvold1}.) 
Our present understanding of the phenomenon of nucleation
is based largely on the work of Langer~\cite{langer}.
In the absence of impurities in the system, Langer's approach is
characterized as homogeneous nucleation theory.
The basic assumption is that the system in the metastable phase is in a state
of quasi-equilibrium. The nucleation rate is proportional to the imaginary part
of the analytic continuation of the equilibrium free energy into the 
metastable phase. 
The rate can be calculated by considering fluctuations of the system 
around particular configurations
characterized as critical bubbles or droplets. They are saddle
points of the free energy and drive the nucleation process. 
Extensive studies have been carried out within this framework
in the last decades. (See the reviews~\cite{abraham,karsch,gunton,binder,rikvold1}, and
ref.~\cite{cott} for work in the context of field theory.)
Also alternative approaches have been pursued, that do not rely on
the explicit introduction of droplets \cite{rikvold2}.

We employ here the application of Langer's approach
to field theory~\cite{coleman, affleck}.
The basic quantity of interest is 
the nucleation rate $I$, which
gives the probability per unit time and volume to nucleate a certain
region of the stable phase (the true vacuum) within the metastable 
phase (the false vacuum). 
The calculation of $I$ relies on an expansion
around a dominant semiclassical saddle-point that is identified with 
the critical bubble or droplet. 
This is a static configuration 
(usually assumed to be spherically symmetric) within the metastable phase 
whose interior consists of the stable phase. 
It has a certain radius that can be determined from the 
parameters of the underlying theory. Bubbles slightly larger 
than the critical one expand rapidly, thus converting the 
metastable phase into the stable one. 

The bubble-nucleation rate is exponentially suppressed by the action
(the free energy rescaled by the temperature)
of the critical bubble.
Possible deformations of the critical  bubble
generate a static pre-exponential factor.
The leading contribution to it
has the form of a ratio of fluctuation determinants and corresponds to the
first-order correction to the semiclassical result. 
Apart from the static prefactor, the nucleation rate 
includes a dynamical prefactor that takes into account the expansion of
bubbles after their nucleation. We concentrate only on
the static aspects of the problem and neglect the dynamical prefactor.

We consider a three-dimensional
statistical system with one space-dependent degree of freedom described
by a real scalar field $\phi(x)$.
For example, $\phi(x)$ may correspond to
the density for the gas/liquid transition,
or to a difference in concentrations for chemical phase transitions,
or to the magnetization for the ferromagnetic transition.
Our discussion also applies to a (3+1)-dimensional quantum field theory in
quasi-thermal equilibrium for energy scales below the temperature.
Then an effective three-dimensional description
applies~\cite{trans}.
In a different context our results can also
be applied to the problem of quantum tunnelling in a (2+1)-dimensional
field theory at zero temperature. 

The bubble-nucleation rate
is given by~\cite{langer,coleman}
\beq
I=A \exp\left(-S_b\right)=\frac{E_0}{2\pi}
\left(\frac{S_b}{2\pi}\right)^{3/2}\left|
\frac{\det'[\delta^2 S/\delta\phi^2]_{\phi=\phibounce}}
{\det[\delta^2 S/\delta\phi^2]_{\phi=0}}\right|^{-1/2}
\exp\left(-S_b\right). 
\label{rate0} \eeq
Here $S$ is the action 
of the system for a given configuration of the
field $\phi$. For statistical systems 
the parameters appearing in $S$ can be related to those
in the Hamiltonian (see~\cite{st2} for an example).
The action of the critical bubble is $S_b
=S\left(\phibounce(r)\right)-S(0)$,
where $\phibounce(r)$ is the spherically-symmetric
bubble configuration and $\phi = 0$ corresponds to the false vacuum. 
The fluctuation determinants are evaluated either 
at $\phi = 0$ or around $\phi=\phibounce(r)$. 
The prime in the fluctuation determinant around
the bubble denotes that the three zero eigenvalues 
of the operator $[\delta^2 S/\delta\phi^2]_{\phi=\phibounce}$
have been removed. 
Their contribution generates the factor 
$\left(S_b/2\pi \right)^{3/2}$ and the volume factor
that is absorbed in the definition of $I$ (nucleation rate per unit volume). 
The quantity $E_0$ is the square root of
the absolute value of the unique negative eigenvalue.

The calculation of the pre-exponential factor $A$ is a technically 
difficult problem. In this letter we compare the results of two different
methods for its evaluation
in order to confirm their reliability. The first method, 
described in refs.~\cite{st1,st2} and denoted by ${\cal A}$ in the 
following, is based on the notion of
coarse graining and employs the Wilson approach to the renormalization
group \cite{wilson} in the formulation of the effective average action
\cite{averact}. It can be applied to a multitude of systems described by
a variety of actions.
In the second method, described in ref.~\cite{mr} and denoted by ${\cal B}$,
the nucleation rate is calculated analytically near the so-called thin-wall
limit. Starting from a bare action with a quartic potential,
the bubble solution, its action, and the fluctuation determinants are
obtained as power-series in an asymmetry parameter, which is small near the
thin-wall limit. Renormalization is performed in the context of dimensional
regularization near 3 dimensions.

\section{Relation between coarse graining and dimensional regularization}
In ref.~\cite{mr} nucleation rates 
were calculated for a model described by the bare 
action\footnote{This model has a special feature:
when the two minima are taken to be
exactly degenerate
a $Z_2$ symmetry guarantees that they are equivalent.
Therefore the thin-wall limit of more generic models 
without this particular feature can be qualitatively different.}
\beq
S_0=\int d^3 x[K_0-V_0]=\int d^3 x\left[
\frac{(\partial \phi)^2}{2}-\frac{m^2}{2}\phi^2 - \frac{\gamma}{6}\phi^3-
\frac{\quartic }{8}\phi^4\right].
\label{bareaction} \eeq
This potential has the typical form relevant for
first-order phase transitions in (3+1)-dimensional
field theories at high temperature.
Through a shift $\phi\to\phi+c$ the cubic term can be eliminated
in favour of a term linear in $\phi$.
As a result the same potential 
can describe statistical systems of the Ising universality class
in the presence of an external magnetic field.
The calculation was performed in the limit that the asymmetry parameter
is small and the two minima of the potential
have nearly
equal depth. The critical bubbles are not far from
the thin-wall limit: The width of the surface is much smaller than 
the radius.

We would like to verify that 
the predictions of method ${\cal A}$ are in agreement 
with those of method ${\cal B}$.
In order to make the comparison we must understand
how to find a coarse-grained action equivalent to a
bare or renormalized one in the framework of dimensional regularization.
This can be done by requiring that the two versions of the same 
theory describe the same physics.
In the one-loop approximation we can require that the same 
effective potential is obtained in both cases.
With dimensional regularization, the one-loop effective potential is
\begin{equation}\label{eq:V1DREG}
V_1 = V_0 - \frac{1}{12\pi}(V''_0)^{3/2}.
\end{equation}
Since poles in $\epsilon=3-d$ only
appear at the two-loop level, in the one-loop approximation the
minimally renormalized parameters are equal to the bare ones.
If desired, they can be related to physical observables
via finite renormalization corrections~\cite{mh,mr}.

\medskip

On the other hand, the coarse graining employed in method ${\cal A}$
can be implemented by introducing an effective infrared
cutoff that acts as a mass term $\sim k^2$ \cite{st1,st2}. 
The coarse-grained potential $V_k$ becomes the effective potential 
for $k=0$. Only fluctuations with
characteristic momenta $0\lta q^2\lta k^2$ contribute to the expression that
relates $V_{k=0}$ and $V_k$:
\begin{equation}
V_1 =
V_{k=0} = V_k - \frac{1}{12\pi}[V_k''{}^{3/2}-(k^2+V''_k)^{3/2}].
\end{equation}
The one-loop effective potential $V_1$ is 
complex for values of $\phi$ such that $V''_k<0$. 
This pathology of the one-loop approximation
could be avoided solving numerically the exact renormalization-group equation for $V_k$.
However, the same problem appears at one loop in the dimensionally regularized version of the theory,
eq.\eq{V1DREG},
and cancels out when matching the two versions of the theory.
Therefore, at 
one-loop order, the coarse-grained potential $V_k$ corresponding
to a given bare
potential $V_0$ employed in dimensional regularization is
\begin{equation}\label{eq:Vk}
V_k = V_0-\frac{1}{12\pi}(k^2+V''_0)^{3/2}.
\label{match} \end{equation}
In the context of method ${\cal A}$
the calculation of nucleation rates is performed at values of $k$ sufficiently 
large that the one-loop approximation for $V_k$, eq.\eq{Vk}, is real for all $\phi$.

In method ${\cal A}$ we 
neglect the corrections to terms with derivatives of the field in the coarse-grained action,
keeping only the corrections to the effective potenatial.
We have verified that including also
the field-dependent one-loop correction to the kinetic term
has a negligible influence on the following results.

\section{Comparison of the two methods}
It is convenient to use the 
rescalings $x=\tilde{x}/m$ and $\phi=2\tilde{\phi}m^2/\gamma$
in order to put $S_0$ in the simplified form 
\beq
S_0=
\frac{4m^3}{\gamma^2}\cdot\int d^3\tilde{x}\left[\frac{(\tilde{\partial} 
\tilde{\phi})^2}{2}-
\frac{\tilde{\phi}^2}{2} - \frac{\tilde{\phi}^3}{3}-
\frac{h}{18}\tilde{\phi}^4\right]=
\frac{1}{\geff}\cdot \tilde{S}.
\label{simp} \eeq
We can now use $h$ and $\geff$ as parameters instead of $m,\gamma,\quartic $.
The dimensionless parameter $h\equiv 9\quartic  m^2/\gamma^2$, 
ranging between $0$ and $1$, controls the shape of the potential:
The barrier is small for small $h$, while
$h=1$ corresponds to two degenerate minima.
Method ${\cal B}$
is valid near the thin-wall limit, i.e.\ for $h$ close to 1.
The dimensionless parameter 
$\geff\equiv \gamma^2/4m^3 = 9\quartic /4mh$ controls the strength of the 
self-interactions of the field, and, therefore, 
the size of the loop corrections.
This can be seen by rewriting eq.~(\ref{match}) as
\beq
\tilde{V}_k = \tilde{V}_0
-\frac{\geff}{12\pi}(\tilde{k}^2+\tilde{V}''_0)^{3/2}
\label{simpl2} \eeq
where $\tilde{k}=k/m$ is a rescaled version of the 
coarse-graining scale $k$.\\
\indent
Before performing the comparison we identify the range of parameters 
$h,g$ where
both approaches give reliable results.
The saddle-point expansion, like any perturbative method,
breaks down if the dimensionless coupling $\geff$ is too large.
In the approach ${\cal A}$
this is signalled by a
strong residual dependence of the nucleation rate on the arbitrary 
coarse-graining scale $k$ at which the calculation is performed. 
The origin of the problem lies in 
the omission of the important higher-order corrections in the
expansion.
In the thin-wall limit, for fixed $\geff$, the relative 
importance of the pre-exponential factor
(that we can quantify as $\left| \ln A /\ln I\right| $, 
with $A$ and $I$ in units of $m$)
is minimal and approximately amounts to a $\geff\cdot 20\%$ correction. 
As a result,
the saddle-point expansion is meaningful for $\geff\circa{<}2$.
The approximation for the saddle-point action employed in
method ${\cal B}$ is valid only for $h$ close to 1.
As a result, the comparison of the two methods can be performed only above a
certain value of $h$. This minimal value of $h$ increases (decreases)
for decreasing (increasing) values of $\geff$. The reason is that 
the pre-exponential factor in the nucleation rate
becomes less (more) important with decreasing (increasing) $\geff$,
so that one needs to start from a more (less) accurate 
approximation for the bubble action in order to have a meaningful comparison.
For example one needs $h\circa{>} 0.8$ when $\geff=1$. 
On the other hand, for very large bubbles
(in practice when $h>0.95$ i.e.\ when $\St\circa{>}10^4$)
it becomes too difficult to 
determine numerically the saddle point action and the pre-exponential factor,
so that method ${\cal A}$ cannot be applied reliably.
In conclusion, there is a range of values 
of $h$ ($0.8<h<0.95$ for $\geff=1$) and of $\geff$ ($\geff\circa{<}2$)
where both
techniques 
can be applied and a comparison is possible.

\begin{rfigure}{6cm}\setlength{\unitlength}{1cm}
\begin{picture}(6,10)
\putps(0.7,0)(0.7,0){fAS}{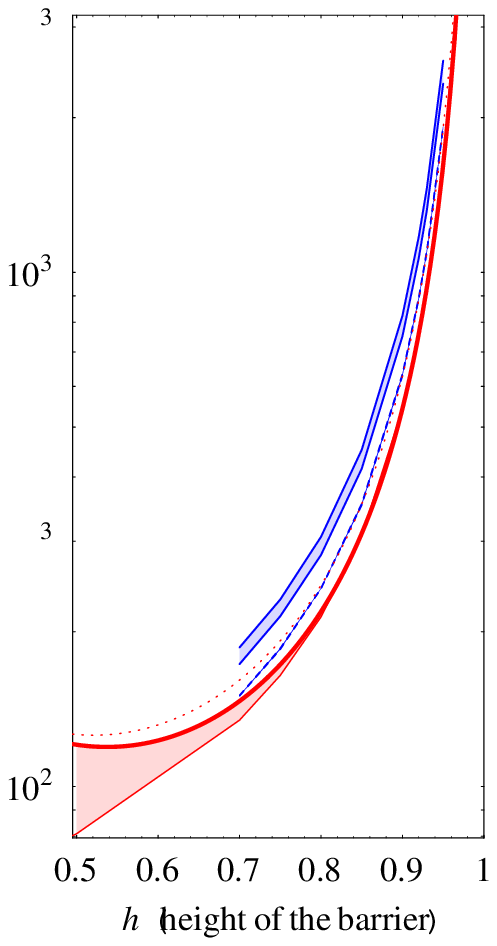}
\end{picture}
\begin{center}
\parbox{5cm}{\caption{\em Values of the 
bubble action and the nucleation rate in the two approaches.
See the text for a description.}}
\end{center}
\end{rfigure}
In fig.~1 we display the bubble action (bands) and 
nucleation rate (lines)
in the two approaches as a function of $h$ for $\geff=1$.
The lower band depicts the value of the
bubble action computed from the bare potential in the approach ${\cal B}$.
The band is delimited by $S_0$, its exact numerical value  (lower thin line),
and by $S_{\cal B}$, the analytical
approximation for it (upper thick line).
The thickness of the band indicates the accuracy of the analytical approximation.
The dotted line corresponds to the prediction of method ${\cal B}$ 
for the nucleation rate $\ln I$, with $I$ in units of $m$.
The upper band depicts the values of the
bubble action $\SAk $ computed from the coarse-grained potential $V_k$
for $m^2 \leq k^2 \leq 2m^2$.
The $k$-dependence of $\SAk $ is compensated by that of the prefactor $A_{{\cal A}k}$.
As a result, the prediction of method ${\cal A}$ for $\ln I$, 
denoted by a continuous line, is $k$-independent.
The overlap of the dotted and continuous lines
indicates that the predictions of the two methods
for the bubble-nucleation rate 
agree when one is not too far from the thin-wall limit.\\
\indent
The region of validity of both approaches
leads to large bubble actions, which makes an accurate
comparison difficult in fig.~1.
For this reason we plot in fig.~2 some useful ratios: 
\renewcommand{\labelenumi}{\theenumi}
\renewcommand{\theenumi}{(\alph{enumi})}
\begin{enumerate}
\item  $(\ln I_{\cal A})/\SAk $ for $m^2 \leq k^2 \leq 2 m^2$
(thick band at the bottom);
\item  $(\ln I_{\cal B})/S_{\cal B}$ (continuous line at the top);
\item  $S_0/S_{\cal B}$ (dashed line);
\item  $(\ln I_{\cal A})/(\ln I_{{\cal B}})$ for $m^2 \leq k^2 \leq 2 m^2$
(thin dark band in the middle),
\item  \parbox[t]{10.2cm}{$(\ln I_{\cal A}^{\rm approx})/(\ln I_{{\cal B}})$ 
for $m^2 \leq k^2 \leq 2 m^2$ (thick light band in the middle), where 
$I_{\cal A}^{\rm approx}$ is obtained using the following
approximation for the prefactor~\cite{st1,st2}}
\begin{equation}
\ln A_{{\cal A}k}^{\rm approx}
\approx \frac{\pi k}{2}\left[- \int_0^\infty r^3 \left[
V''_k\left( \phi_{\rm b}(r) \right)-V''_k\left( 0 
\right)\right] dr \right]^{1/2}.
\label{eq:appr}
\end{equation}
\end{enumerate}
{}From the form of these ratios the following conclusions can be reached:
\begin{itemize}
\item 
Band (a) shows that the prefactor gives a significant contribution
to the total nucleation rate $\ln I_{\cal A}$ in approach ${\cal A}$. 
The strong $k$ dependence of this ratio is
due to $\SAk $. The rate is $k$-independent 
to a very good approximation for $\geff \lta 1$.
\item 
Line (b) demonstrates that the prefactor is important 
also in approach ${\cal B}$.  
\item 
Line (c) indicates the region of validity of the expansion around the
thin-wall approximation
employed in approach ${\cal B}$. The analytical expression for the
bubble action $S_{\cal B}$ gives a good approximation to the numerically 
computed $S_0$ only 
above a certain value of $h$ that
increases with decreasing $\geff$.
\item 
The small width of band (d) indicates that the $k$ dependence of
$\ln I_{\cal A}$ is very weak. This verifies that approach ${\cal A}$
is reliable. 
The fact that 
band {\rm (d)} is close to one, {\em  demonstrates 
that the two approaches agree very well when they are both reliable}.
\item
Away from the thin-wall limit, $S_{\cal B}$  
overestimates the true bubble action $S_0$.
At $h\approx 0.7$, for example, $S_{\cal B}$ is $\sim10\%$ larger than $S_0$.
For $\geff \lta 1$ the approximated prefactor is small, $|\ln A_{\cal B}|\circa{<}10\% \cdot S_{\cal B}$,
and presumably  has a only a $\sim10\%$ error.
As a result, method ${\cal B}$
is not sufficiently accurate and the two approaches do not agree.
However, the fact that line (c) and band (d) deviate from 1 in exactly the
same way indicates that the disagreement is largely caused by the inaccurate 
determination of $S_0$, while the estimates of the prefactors in the
two approaches are in agreement.
\item
Bands (d) and (e) are in good agreement below the value of $h$ at which
the approach ${\cal B}$ stops being reliable. This makes it possible to
extend the region of $h$ for which an analytical expression for the
nucleation rate is available.
Near  the thin wall limit ($h$ close to 1), the analytical expression provided by 
method ${\cal B}$ (eq.~(138) or eq.~(147) of ref.~\cite{mr}) can be
used. For smaller $h$, when the  expansion around the
thin-wall approximation breaks down
(as can be checked by comparing eq.~(54) of ref.~\cite{mr} 
with a numerical determination of the bubble action) eq.~(\ref{eq:appr})
can be used.
\end{itemize}
\begin{figure}[t]
\begin{center}\setlength{\unitlength}{1cm}
\begin{picture}(19,5.5)
\putps(-1.3,0)(-1.3,0){fcp}{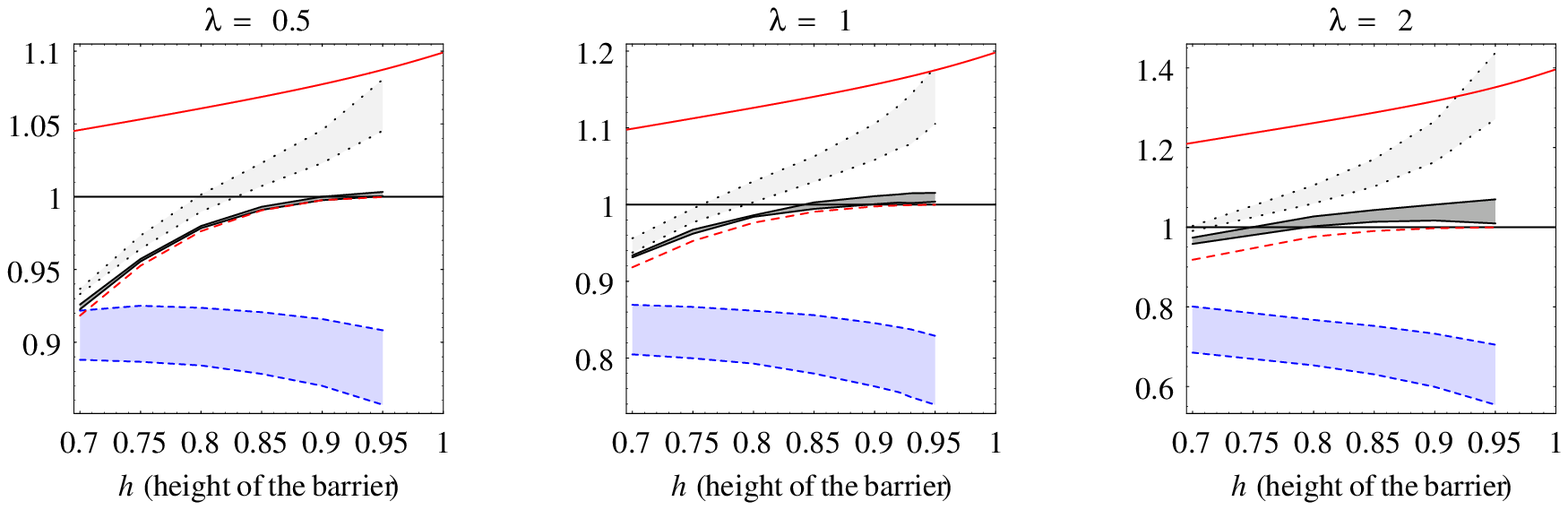}
\put(16.5,1.5){\Blue{\tiny (a)}\Black}
\put(13.2,4.5){\Red{\tiny (b)}\Black}
\put(13.2,2.7){\Red{\tiny (c)}\Black}
\put(16.5,3.4){{\tiny (d)}}
\put(16.5,4.5){{\tiny (e)}}
\end{picture}
\caption[SP]{\em Comparison of the two methods
for three different values of the coupling $\geff$.
The continuous line {\rm (b)} at the top of each plot indicates 
the importance of the prefactor in the 
approach ${\cal B}$, estimated through
$(\ln I_{\cal B})/S_{\cal B}$.
The thick band at the bottom {\rm (a)} indicates the 
importance of the prefactor in the approach ${\cal A}$,
estimated as 
$(\ln I_{\cal A})/\SAk $ and computed for $m^2\leq k^2 \leq 2m^2$.
The thin dark band in the middle {\rm (d)} 
depicts the ratio of the nucleation rates 
$(\ln I_{\cal A})/(\ln I_{{\cal B}})$
computed in the two approaches.
The light band {\rm (e)}  depicts 
$(\ln I_{\cal A}^{\rm approx})/(\ln I_{{\cal B}})$,
where $\ln I_{\cal A}^{\rm approx}$ is obtained using the approximation in eq.\eq{appr}.
The dashed line {\rm (c)}  corresponds to $S_0/S_{\cal B}$ and indicates
how much the approximations of method ${\cal B}$
overestimate the true saddle-point action.
All dimensionful quantities are in units of $m$.
\label{fig:Res}}
\end{center}\end{figure}

Finally, we point out that in
approach ${\cal A}$ fluctuations around the
bubble (taken into account by the prefactor)
enhance the nucleation rate with respect to $\exp(-\SAk )$.
On the other hand,
in approach ${\cal B}$ 
the nucleation rate is smaller than its tree-level approximation,
when the comparison is performed at fixed values of the bare parameters.
This does not indicate a discrepancy.
The prefactor enhances the nucleation rate in approach ${\cal B}$ as well,
if the comparison
is done at fixed values of the physical mass and couplings defined at the true minimum~\cite{mr}.

\section{Conclusions}
In their joint region of validity the two methods for the calculation of
the nucleation rate, based on coarse-grained potentials on the one hand and
on dimensional regularization on the other, agree very well, thus
giving support to their reliability.

\paragraph{Acknowledgments}
The work of N.T. was supported by the E.C. under TMR contract 
No. ERBFMRX--CT96--0090 and contract No. ERBFMBICT983132.

\end{document}